\documentclass[fleqn,usenatbib]{mnras}
\usepackage{newtxtext,newtxmath}

\usepackage[T1]{fontenc}
\usepackage{ae,aecompl}
\usepackage{times}
\usepackage{rotating}
\usepackage{graphicx}	
\usepackage{amsmath,amstext}	

\usepackage{amssymb}	

\usepackage{soul}

\newcommand{\Msun}{\,{\rm M_{\odot}}}

\newcommand{\epsff}{\epsilon_{\mathrm{ff}}}

\title[GMCs in simulated galaxies]{The effects of subgrid models on the properties of giant molecular clouds in galaxy formation simulations}

\author[Hui~Li et al.]
    {\parbox[T]{18cm}{Hui Li$^{1}$\href{http://orcid.org/0000-0002-1253-2763}{\includegraphics[scale=0.8]{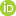}}\thanks{E-mail: hliastro@mit.edu}\thanks{NHFP Hubble Fellow}
    , Mark Vogelsberger$^{1}$\href{http://orcid.org/0000-0001-8593-7692}{\includegraphics[scale=0.8]{figure/orcid.png}},
    Federico Marinacci$^{2}$\href{http://orcid.org/0000-0003-3816-7028}{\includegraphics[scale=0.8]{figure/orcid.png}},
    Laura V. Sales$^{3}$\href{https://orcid.org/0000-0002-3790-720X}{\includegraphics[scale=0.8]{figure/orcid.png}},
    Paul Torrey$^{4}$\href{http://orcid.org/0000-0002-5653-0786}{\includegraphics[scale=0.8]{figure/orcid.png}}
     }\vspace{0.2cm}\\
     $^{1}$Department of Physics and Kavli Institute for Astrophysics and Space Research, Massachusetts Institute of Technology, Cambridge, MA 02139, USA \\
     $^{2}$Department of Physics \& Astronomy, University of Bologna, via Gobetti 93/2, 40129 Bologna, Italy\\
     $^{3}$Department of Physics \& Astronomy, University of California, Riverside, 900 University Avenue, Riverside, CA 92521, USA\\
     $^{4}$Department of Astronomy, University of Florida, 211 Bryant Space Sciences Center, Gainesville, FL 32611, USA\\
    }

\date{Accepted XXX. Received YYY; in original form ZZZ}

\pubyear{2019}

\begin{document}
\label{firstpage}
\pagerange{\pageref{firstpage}--\pageref{lastpage}}
\maketitle

\begin{abstract}
Recent cosmological hydrodynamical simulations are able to reproduce numerous statistical properties of galaxies that are consistent with observational data. Yet, the adopted subgrid models strongly affect the simulation outcomes, limiting the predictive power of these simulations. In this work, we perform a suite of isolated galactic disk simulations under the {\it SMUGGLE} framework and investigate how different subgrid models affect the properties of giant molecular clouds (GMCs). We employ {\sc astrodendro}, a hierarchical clump-finding algorithm, to identify GMCs in the simulations. We find that different choices of subgrid star formation efficiency, $\epsff$, and stellar feedback channels, yield dramatically different mass and spatial distributions for the GMC populations. Without feedback, the mass function of GMCs has a shallower power-law slope and extends to higher mass ranges compared to runs with feedback. Moreover, higher $\epsff$ results in faster molecular gas consumption and steeper mass function slopes. Feedback also suppresses power in the two-point correlation function (TPCF) of the spatial distribution of GMCs. Specifically, radiative feedback strongly reduces the TPCF on scales below 0.2~kpc, while supernova feedback reduces power on scales above 0.2~kpc. Finally, runs with higher $\epsff$ exhibit a higher TPCF than runs with lower $\epsff$, because the dense gas is depleted more efficiently thereby facilitating the formation of well-structured supernova bubbles. We argue that comparing simulated and observed GMC populations can help better constrain subgrid models in the next-generation of galaxy formation simulations.
\end{abstract}

\begin{keywords}
galaxies: evolution -- galaxies: ISM -- galaxies: structure -- ISM: clouds -- methods: numerical
\end{keywords}



\section{Introduction}\label{sec:intro}

During the last few decades, cosmological hydrodynamical simulations have become one of the most powerful tools to study the formation and evolution of galaxies \citep[see][for a recent review]{vogelsberger_etal19}.
Due to ever increasing computing power, both large-scale \citep[e.g.][]{vogelsberger_etal14,dubois_etal14,schaye_etal15,davee_16,tremmel_etal17,springel_etal18, dave_etal19} and zoom-in \citep[e.g.][]{guedes_etal11,hopkins_etal14,wang_etal15,wetzel_etal16,grand_etal17,henden_etal18} cosmological simulations have made rapid progress in reproducing various types of galaxies and several galactic scaling relations.
This success is partly due to more accurate treatments of complex astrophysical processes, such as gravity, gas dynamics, and radiative heating and cooling. However, the most significant advances have been a result of novel implementations that describe physical processes that cannot be spatially or temporally resolved in these simulations, such as star formation \citep[e.g.][]{cen_ostriker_92,katz92,navarro_white_93,springel_hernquist03, li_etal17} and stellar feedback \citep[e.g.][]{stinson_etal06,agertz_etal13, vogelsberger_etal13, ceverino_etal14, hopkins_etal14, smith_etal18, marinacci_etal19} subgrid models.

These subgrid models are typically calibrated by the observed properties of galaxies, such as the galaxy luminosity functions, star formation histories (SFHs), and Kennicutt-Schmidt relation.
However, it is unsettling that simulations with disparate and sometimes contradictory subgrid models produce galaxies with similar global properties by fine-tuning their model parameters, thus reducing their predictive power \citep[][]{naab_ostriker17}.
Moreover, even though these simulations reproduce many galactic properties, it is still unknown whether they capture the small-scale structures of the interstellar medium (ISM) properly.
As the spatial and mass resolutions of the simulations, especially the zoom-in ones, are approaching the size and mass of individual star-forming regions \citep[e.g.][]{wetzel_etal16,li_etal17,hopkins_etal18,wheeler_etal19,agertz_etal19,lahen_etal19}, it is urgent to test these subgrid models using observables on similar scales.

Most stars in galaxies are formed in star clusters, which emerge from cold and dense giant molecular clouds (GMCs). The observed linear correlation between molecular gas and star formation surface density \citep[e.g.][]{kennicutt98, gao_solomon04, genzel_etal10,garcia-burillo_etal12, tacconi_etal13} suggests that molecular gas is a direct indicator of star formation activity in galaxies. 
Over the past few decades, many observational studies have already been conducted to systematically investigate the statistical properties of GMCs in both the Milky Way \citep[see a recent review][]{heyer_dame15} 
and nearby face-on galaxies \citep[e.g.][]{engargiola_etal03,rosolowsky05,rosolowsky_blitz05,hughes_etal13,meidt_etal13}.
Most recent sub-mm surveys of nearby galaxies, such as PHANGS-ALMA \citep[e.g.][]{sun_etal18} and ALMA-LEGUS \citep{grasha_etal18,grasha_etal19}, have delivered detailed molecular gas distributions with unprecedented resolution and sensitivity that can resolve individual GMCs of mass $\sim10^4\Msun$ and size $\sim10$~pc, coincident with the resolutions of the most recent galaxy formation simulations.
Comparing the simulations with these sub-mm observations offers a great opportunity to improve subgrid implementations in cosmological simulations that aim at resolving the multi-phase ISM and star-forming regions.

Indeed, different types of galaxy formation simulations have investigated the properties of GMCs, such as mass function, velocity dispersion, and virial parameter, during the part decade. Previous isolated galactic disk simulations focused on how galactic environments affect the evolution of GMCs and how galactic shears and cloud-cloud collisions provide turbulent energy for individual clouds \citep[e.g.][]{tasker_tan09, tasker11, dobbs_etal11, dobbs_pringle13, ward_etal16}. Moreover, the inclusion of molecular chemistry in the simulations was also shown to be an important factor for direct comparison with sub-mm observations \citep[e.g.][]{khoperskov_etal16, duarte-cabral_dobbs17, nickerson_etal19}. However, most of these simulations did not include localized stellar feedback from multiple feedback channels that help disrupt GMCs and maintain a multiphase ISM.
These feedback processes are demonstrated to be essential to the prediction of the ISM structure and cloud properties \citep[e.g.][]{hopkins_etal12, grisdale_etal18, benincasa_etal19}.
On the other hand, several recent cosmological simulations have explored the effects of star formation and stellar feedback subgrid models on the properties of molecular gas and star clusters.
In \citet{li_etal18}, we investigated how the subgrid star formation prescriptions affect the various properties of young massive clusters. We showed that the initial cluster mass function and the cluster formation efficiency depends strongly on the choice of the local star formation efficiency per free-fall time, $\epsff$. Using a subset of NIHAO simulations, \citet{buck_etal19} showed that the choice of star formation density threshold changes the spatial clustering of young stars and favours a high values for the threshold.

To systematically investigate the effects of subgrid models on the properties of molecular gas in galaxy formation simulations, in this paper, we perform a suite of high resolution simulations of isolated Milky Way-sized galaxies using the Stars and MUltiphase Gas in GaLaxiEs -- {\it SMUGGLE} model, an explicit and comprehensive stellar feedback framework \citep[][M19]{marinacci_etal19} for the moving-mesh code {\sc arepo} \citep{springel10arepo}. We run the simulations starting from the exact same initial conditions but with different variations of subgrid models and parameters. We study the mass and spatial distribution of GMCs as identified by the hierarchical clump-finding algorithm, {\sc astrodendro}. Our paper is organized as follows. In \autoref{sec:methods}, we briefly summarize the physical processes involved in the {\it SMUGGLE} model, describe the setup of different model variations, and illustrate the workflow to identify GMCs from the simulation snapshots. In \autoref{sec:results}, we show the similarity and differences of the various properties of the molecular gas, such as the gas density profile, mass function, and two-point correlation function (TPCF) of the model GMCs, for different model variations. In \autoref{sec:discussion}, we compare our results with some previous studies and discuss a few caveats in our numerical experiments. Finally, we summarize the key results of the paper in \autoref{sec:summary}.

\begin{figure*}
\includegraphics[width=2.0\columnwidth]{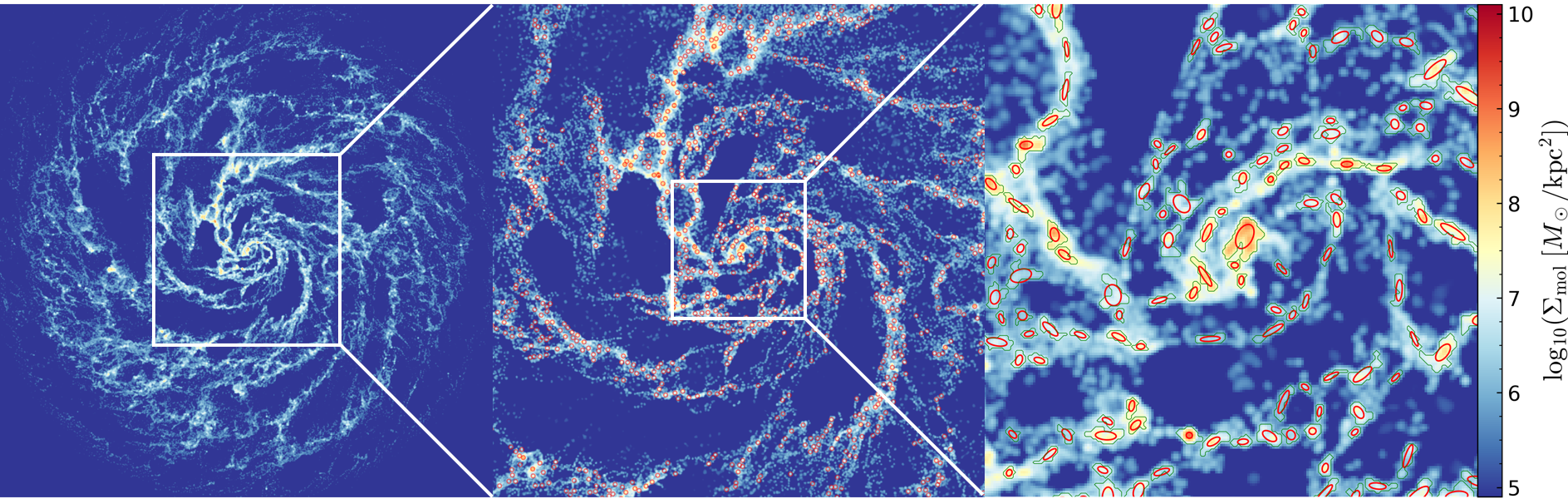}
\vspace{0mm}
\caption{Identification of GMC candidates from the molecular surface density projection using the {\sc astrodendro} algorithm. \textit{Left:} Face-on projection of molecular surface density for the whole galactic disk in SFE1 run at 0.5~Gyr. \textit{Middle:} Spatial distribution of the identified GMCs indicated with red circles. \textit{Right:} The identified GMCs are shown as green iso-density contours, while the best-fit ellipse of each clump is overplotted as red ellipse. The physical size of these three plots are 40, 16, and 4~kpc on a side, respectively. All panels share the same surface density scale as shown in the colorbar on the right hand side.}
  \label{fig:prj}
\end{figure*}

\section{Methods}\label{sec:methods}

In this section, we first recap some key physical ingredients of the {\it SMUGGLE} model and provide the information on the initial conditions, simulation resolutions, and different variations of subgrid star formation and stellar feedback models. We also describe the analysis procedure for identifying GMCs from the simulation snapshots using the hierarchical clump-finding algorithm, {\sc astrodendro}.

\subsection{Isolated Milky Way-sized galaxy with {\it SMUGGLE}}\label{sec:methods-ISM}

The simulations presented in this paper are performed with the moving-mesh finite-volume hydrodynamic code {\sc arepo}. In {\sc arepo}, the control volumes are discretized by a Voronoi tessellation, which is generated from its dual Delaunay tessellation determined by a set of mesh-generating points. Our simulations include hydrodynamics, self-gravity, radiative heating/cooling, star formation, and stellar feedback using the {\it SMUGGLE} model. The model incorporates explicit gas cooling and heating over a large range of temperatures between 10--$10^8$~K so that the thermodynamical property of the ISM is modeled explicitly.

More importantly, {\it SMUGGLE} adopts physically-motivated star formation and feedback subgrid models. Star particles are formed from cold, dense, and self-gravitating molecular gas at a rate that depends on the star formation efficiency per free-fall time $\dot{M}_* = \epsff M_{\rm gas}/\tau_{\rm ff}$,
where $\dot{M}_*$, $M_{\rm gas}$, and $\tau_{\rm ff}$ are the star formation rate, gas mass, and free-fall timescale of a given Voronoi cell above the star formation density threshold, $n_{\rm th}=100/\rm cm^3$.
All relevant stellar feedback processes, such as photoionization, radiation pressure, energy and momentum injection from stellar winds and supernovae (SNe), are included. Because each star particle in the simulations represents a single stellar population, the mass, momentum, and energy deposition rates are determined by the IMF-averaged values assuming a \citet{chabrier03} IMF. SN events are sampled discretely via a Poisson process with a rate determined by the IMF and SN progenitor mass. Unless the cooling radius of the supernova remnants is resolved locally, the $PdV$ work in the Sedov-Taylor phase is modeled via direct momentum deposition. Feedback energy and momentum from young star particles are deposited to 64 nearest gas cells in a solid angle-weighted fashion. Momentum feedback of fast and slow winds from OB and AGB stars is deposited in a similar fashion. While the simulations do not include direct treatment of radiative transfer, we treat photoionization around young star particles by imposing a temperature floor of $1.7\times10^4$~K to the nearby gas cells stochastically based on the total budget of ionizing photons. We refer the reader to the flagship {\it SMUGGLE} paper, for a detailed description of the numerical implementation. This model successfully reproduces the multiphase ISM structure, generates galactic fountain flows self-consistently, and maintains feedback-regulated inefficient star formation that is consistent with observations.

The initial conditions used in this paper are the same as that of M19.
It contains a Milky Way-sized galaxy of total mass of $1.6\times10^{12}\Msun$, which consists of a stellar bulge and disc, a gaseous disc, and a dark matter halo, whose masses are similar to the Milky Way. The gaseous disc has a total mass of $\approx 9\times10^9\Msun$ and initially has an exponential profile with a scale length of 6~kpc. This initial setup gives a gas fraction around 10\% within $R_\odot=8.5$~kpc. In order to resolve molecular clouds more massive than $\sim10^4\Msun$, the mass resolution of the simulation is around $1.4\times 10^3 \Msun$ per gas cell, which corresponds to that of the highest resolution runs in M19. The gravitational softening for gas cells is adaptive, with a minimum softening of about 3.6~pc for all simulations presented in this paper.

\begin{table}
 \label{tab:models}
\centering
 \begin{tabular}{c c c c} 
 \hline
    Name  & $\epsff$ & SN & Radiative  \\
 \hline
    SFE1  & 0.01      & Yes& Yes \\
    SFE10  & 0.1      & Yes& Yes \\
    SFE100  & 1.0      & Yes& Yes \\
    Nofeed & 0.01      & No & No \\
    SN  & 0.01      & Yes& No \\
    Rad  & 0.01      & No& Yes \\
 \hline
 \end{tabular}
 \caption{Summary of the six model variations in this paper. The column ``SN'' means the SN energy and momentum feedback, while ``Radiative'' means both the photoionization and radiation pressure from young stars.}
\end{table}
\subsection{Model variations}\label{sec:methods-models}

Starting from the same initial conditions, we vary the star formation and stellar feedback subgrid models to investigate how they affect the properties of the GMCs in the isolated Milky Way-sized galaxy. We change the value of $\epsff$ from 0.01 to 1 to study how the rate of local star formation efficiency affects both the properties of the galaxy and individual star-forming regions. We switch on and off different stellar feedback channels, e.g. SN and radiative feedback, to study how different forms of feedback influence the properties of molecular gas on different scales. Below, we describe the six different model variations in detail and list the key parameters in \autoref{tab:models}. For other model parameters that are not mentioned here, we use the same values as the fiducial run (High) in M19.

\begin{itemize}
    \item ``SFE1'': fiducial run (High) in M19 with $\epsff=0.01$ and full suite of stellar feedback mechanism.
    \item ``SFE10'': the same as ``SFE1'' except with $\epsff=0.1$.
    \item ``SFE100'': the same as ``SFE1'' except with $\epsff=1$.
    \item ``Nofeed'': the same as ``SFE1'' but with no stellar feedback.
    \item ``Rad'': the same as ``SFE1'' but with only radiative feedback (photoionization and radiative pressure).
    \item ``SN'': the same as ``SFE1'' but with only SN feedback.
\end{itemize}

\subsection{GMC Identification with {\sc astrodendro}}\label{sec:methods-dendro}

To quantify the effects of subgrid models on the properties of GMCs, we need to establish a robust way to identify GMCs from the simulation output.
We adopt the methodology used by previous observations and focus on the GMC identification in 2D molecular surface density maps. 

We first generate the molecular gas surface density projection along the $z$-axis for a given snapshot, corresponding to a face-on orientation in our simulations. The molecular gas density for each cell is estimated based on the prescription of \citet{mckee_krumholz10} and \citet{krumholz_gnedin11}, where the molecular fraction depends on the metallicity and gas surface density. We focus on the central $40\times40$~kpc region of the galaxy and resolve it with $10\,000\times10\,000$ pixels, therefore the size of each pixel in the projection map is $4\times4$~pc. To better mimic the resolution of recent ALMA observations, we smooth the projection maps with a Gaussian kernel\footnote{We vary the size of the Gaussian kernel between 4 and 16~pc and find that the density projection and the properties of the identified GMCs are not sensitive to the choice of the kernel size.} of similar size as the ALMA beam used in observations \citep[][]{grasha_etal18}. As an example, the left panel of \autoref{fig:prj} shows one representative projection map after smoothing for SFE1 run at 0.5~Gyr.

After the molecular gas surface density map is prepared, we identify dense structures as GMCs using {\sc astrodendro}, a dendrogram-based clump-finding algorithm that has been used extensively to identify molecular clumps in observations for various astronomical purposes, e.g. finding molecular cores in star-forming regions and GMCs in different galaxies \citep{rosolowsky_etal08}. Different from other clump-finding tools, {\sc astrodendro} is based on local segmentation and reveals the hierarchical relationship among clumps and is demonstrated to identify reliable structures in both position-position and position-position-velocity data structures (for a comparison of different clump-finding algorithms, see \citealt{li_clump19}).

The same as other clump-finding algorithms, {\sc astrodendro} requires a few parameters to define the boundaries of structures of interests: min\_value ($\sigma_{\rm base, min}$, the minimum value of the field to be considered as an overdensity), min\_delta ($\sigma_{\rm delta, min}$, minimum significance for structures to avoid including small local maxima caused by fluctuation), and min\_npix ($N_{\rm pix, min}$, the minimum number of pixels of a clump to be identified).
How to choose the values of these parameters is somewhat subjective. To better compare our simulation results with observations, we decide to use the values that are adopted from observations as a guideline. When analyzing observational data with {\sc astrodendro}, $\sigma_{\rm base, min}$ corresponds to the minimum signal-to-noise ratio above which the clumps are considered. The noise level of observation depends on the sensitivity of the telescope. In \citet{grasha_etal18}, they reported a noise level around $\rm I_{CO, noise}^{2-1} \sim 4mJy/beam$ with a velocity resolution of 1.2~km/s. Giving their angular resolution of 0.85 arcsec, a conversion factor between CO(1-0) and CO(2-1) \citep[e.g.][]{ sakamoto_etal99,sawada_etal01}, and an X-factor \citep[e.g.][]{solomon_etal83,dickman_etal86}, the corresponding noise level of the molecular gas surface density is $\sigma_{\rm ALMA} \rm \sim 9.2\times10^5 \Msun/kpc^2$. We use $\sigma_{\rm base, min}=18.4\times10^5 \Msun/\rm kpc^2$ as our fiducial value, which basically means a $2\sigma_{\rm ALMA} $ significance. Moreover, we used $\rm \sigma_{delta, min}=9.2\times10^5 \Msun/kpc^2$ ($1 \sigma_{\rm ALMA} $) and $N_{\rm pix, min}=64$ to avoid any insignificant and unresolved clumps, similar to the values used in a few ALMA observations. To systematically investigate the sensitivity of the properties of the identified GMCs to the choice of all three parameters, we vary each parameter around the fiducial values and compare the mass function of GMCs in different combination of the three parameters. We find that varying the three parameters only changes the distribution of GMCs at the low-mass end (e.g. $<10^4\Msun$) but does not change the results much for more massive GMCs, see Appendix \ref{sec:appendix} for details. As the low-mass GMCs are close to the resolution limit of the simulations, for all quantitative analysis below, we simply discard all GMCs less massive than $10^4\Msun$.

The middle panel of \autoref{fig:prj} shows the distribution of the identified GMCs run using the fiducial {\sc astrodendro} parameters described above. We find that {\sc astrodendro} does an excellent job on identifying GMCs in different environments, such as in the galaxy nuclear, spiral arms, and inter-arm regions. In the right panel, we show the molecular gas distribution for the inner $4\times4$~kpc region of the galaxy together with the isodensity contours of individual clumps and the best-fit ellipses that represent the first and second moments of the clumps. As expected, most of the GMCs follow the spiral structure of the host galaxies; some of them are actually beads along the rings of superbubbles. Most of the GMCs are not spherical in shape. The major axis of the best-fit ellipses largely follows the orientation of large-scale filamentary structures, suggesting that dense gas is stretched along the spiral arms and is compressed in perpendicular directions. As we will see later, the properties of the identified GMC populations depend strongly on the choice of subgrid models in the simulations.

\begin{figure}
\includegraphics[width=1.0\columnwidth]{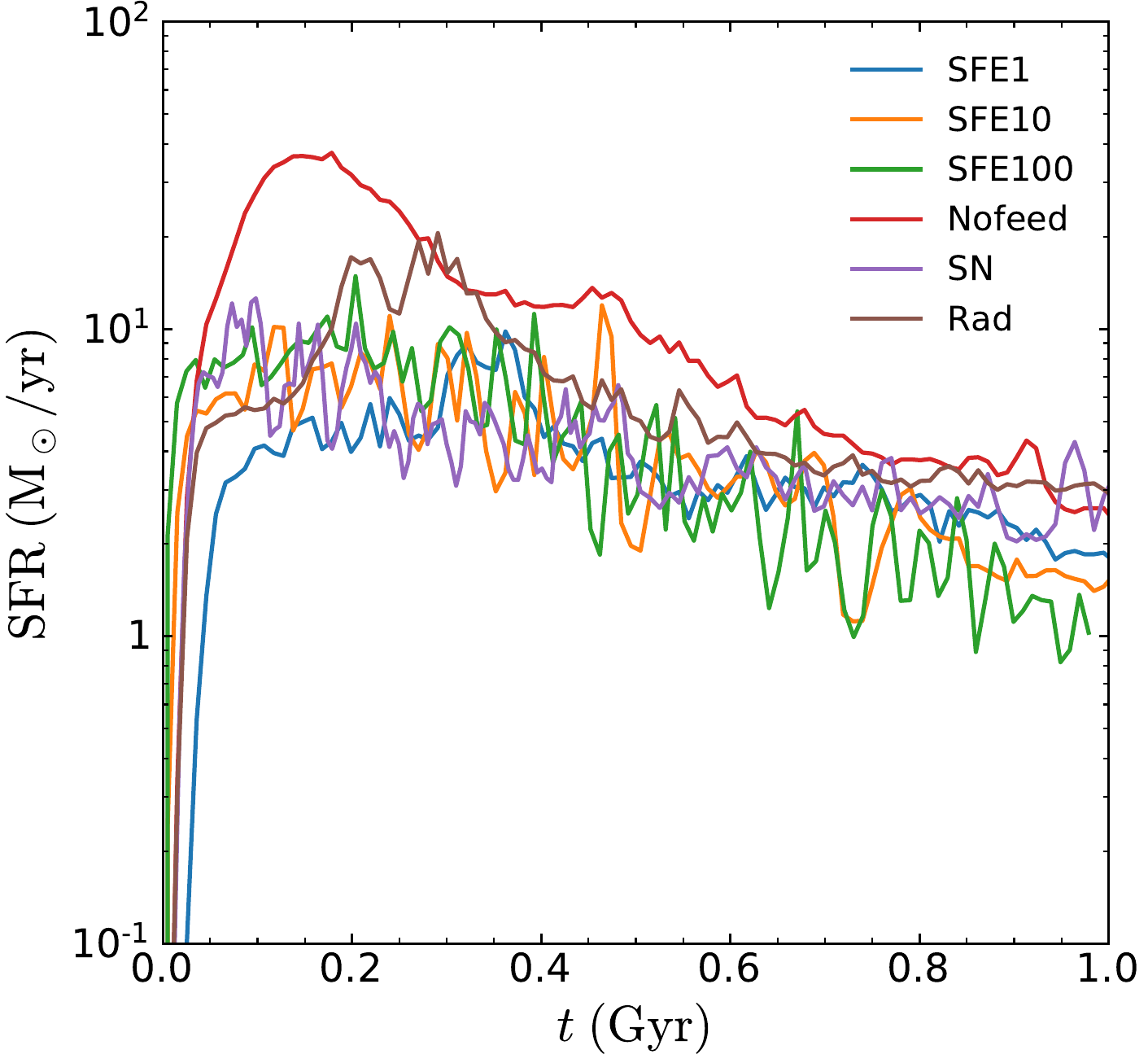}
\vspace{0mm}
\caption{Star formation histories (SFHs) of the simulated galaxy during the first 1~Gyr in all model variations: SFE1 (blue), SFE10 (yellow), SFE100 (green), Nofeed (red), SN (purple), and Rad (brown). The details of the model variations are described in \autoref{sec:methods-models}.}
  \label{fig:sfh}
\end{figure}

\section{Results}\label{sec:results}
We run the simulations from the same initial conditions of an isolated Milky Way-sized galaxy with all six model variations described in \autoref{sec:methods-models}. All simulations are run for 1~Gyr. Since the gas disc needs sometime to relax and settle down to a new equilibrium configuration, we analyze all properties of GMCs after 0.4~Gyr when the multiphase ISM is fully developed with the help of initial star formation and stellar feedback activities.
As can be seen below, after this epoch the star formation rate becomes fairly stable for most of the runs. We utilize the clump-finding algorithm, which is described in \autoref{sec:methods-dendro}, on simulation snapshots between 0.4 and 1~Gyr with a separation of 50~Myr for each run and quantify the variation of the GMC properties over a long period of simulation time. Although the actually snapshot storage frequency is much higher (every 1~Myr), we choose this 50~Myr separation for analyzing GMC properties to avoid identifying the same GMCs of different evolution stages across multiple snapshots.

\subsection{Star formation histories}\label{sec:results-SFH}
Before investigating the properties of GMCs, we first examine the changes of SFH for different model variations.
\autoref{fig:sfh} shows the SFHs of the Milky Way-sized galaxy in six model variations during the first 1~Gyr. The star formation rate (SFR) is averaged over 10~Myr so that the stochasticity caused by individual star-forming regions is smoothed out while the galactic-scale variation is captured.

Without feedback, as expected, the Nofeed run reaches a much higher SFR, $>30\Msun$/yr, after the first dynamical time of the simulation due to the initial gravitational runaway collapse. The SFR gradually decreases after 0.2~Gyr because the initial star burst consumes a large fraction of gas mass in the galactic disk. Although the fast decreasing gas mass, the SFR in the Nofeed run is still always higher than all other runs with stellar feedback throughout the whole simulation period.
In contrast, runs with stellar feedback shows a fairly low SFR around a few $\Msun/$yr, consistent with the current value of the Milky Way, although SFR declines slightly as a function of time. This decline is likely due to gas consumption via star formation and no replenishment of gas from accretion because of our idealized setup. As described in the above section, we identify GMCs in many snapshots between 0.4 and 1 Gyr and report the summary statistics of GMC properties to minimize the stochasticity from individual snapshots. For runs with higher $\epsff$, the SFR rises more dramatically during the first 0.2~Gyr. This is because the conversion from gas to star is faster with higher $\epsff$ and, because the galaxy is still settling down to a new equilibrium, stellar feedback does not have enough time to react and regulate star formation. After 0.4~Gyr, runs with different $\epsff$ have similar average SFR, although higher $\epsff$ runs exhibit more burstiness. The SN run has a significantly higher SFR during the first few hundred Myr, but gradually reach a stable SFR that is similar to the SFE1 run. The Rad run shows similar behavior, but with a SFR always slightly higher than than other runs with SN feedback, though the SFR in Rad is still in a reasonable range for a Milky Way-sized galaxy.

\begin{figure*}
\includegraphics[width=2.0\columnwidth]{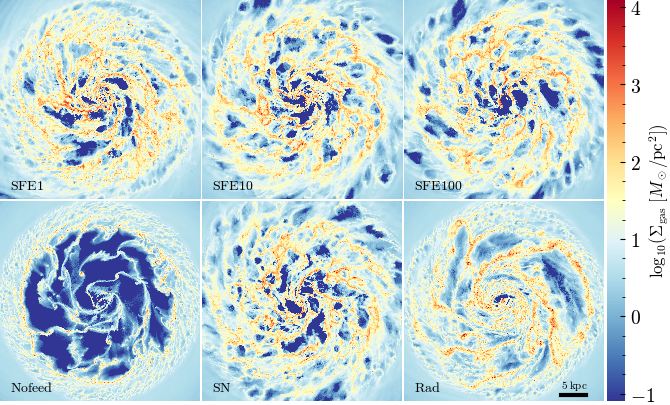}
\vspace{0mm}
\caption{Face-on view of the gas surface density projection at 0.5~Gyr for all model variations, whose names are labeled at the lower left corner of each panel. The physical size of each projection is $40\times40$~kpc.}
  \label{fig:prj-gas-runs}
\end{figure*}

\subsection{Visual impression of the gaseous disk in different runs}\label{sec:results-prj}

As discussed in the previous section, with the exception of the Nofeed simulation the galactic SFR is not dramatically different between the runs. This means that the global star formation activity is not very sensitive to the choice of different subgrid models/parameters for Milky Way-sized galaxies. We now investigate the properties of the ISM and examine the similarities and differences for different model variations. Before presenting the quantitative results, we first give some visual impression of these simulations and describe the characteristic features of the gas distribution qualitatively.

\autoref{fig:prj-gas-runs} shows the gas surface density projected along the $z$-axis (face-on view) at 0.5~Gyr when all model variations reach a roughly stable SFR. In all runs, the dense gas tends to distribute along the spiral structures as well as the central region of the galaxy. We find that, although the galactic SFR for all runs (except Nofeed) is similar at this epoch, the gas distribution differs strikingly.

In SFE1, the gas disk is well-structured, consists of dense gas clumps along the spiral arms and many low density, high temperature cavities that are created by SN explosion from young massive star clusters. The spiral arms are not persistent but short-lived, and are constantly disrupted by various feedback processes launched from dense star-forming regions. The feedback bubbles compress the ambient ISM and trigger the formation of dense gas clumps along the edge of the bubbles.

In contrast, in the Nofeed run, a large fraction of gas mass is concentrated in several large and massive gas clumps. These clumps orbit around the galactic disk, accrete gas mass, and carve out kpc-scale low density regions. Because there is no stellar feedback to terminate the gas accumulation onto these clumps, the mass of these clumps is only limited by star formation and the large-scale galactic shears.

Compared to SFE1, the Rad run does not show many low density cavities. Instead, the gas disk is dominated by several prominent and long-lasting spiral arms where massive GMCs are located. The lack of cavities and the existence of long spiral arms suggest that radiative feedback alone is not able to create kpc-scale superbubbles and puncture holes through the galactic disk vertically. The suppression of star formation happens locally within the scales of individual star-forming regions, where dense gas is heated and dispersed by photoionization and radiation pressure.

The gas surface density distribution for runs with different $\epsff$, SFE1, SFE10, and SFE100, is quite similar, but there are some subtle differences. For example, in SFE100 run, dense gas is organized into many shell-like structures of similar sizes along the spiral arms. These shells are well-arranged across the whole disk and most of the dense gas is distributed along the edge of the bubbles, where it is compressed by shocks from different directions.
The regularity of the shell structure in SFE100 is not seen in SFE1. Instead, in SFE1 run, low-density bubbles have dramatically different sizes and are much less-organized.
The possible explanation of this difference is the following. Higher $\epsff$ leads to faster gas consumption and shorter lifetime of dense gas. Therefore, the gas density distribution cuts off to much lower density for higher $\epsff$. This means the environment that SN feedback acts onto has narrower dynamical range for the gas density. The similarity of the density of star-forming regions leads to a similar size of the SN-driven bubbles. In contrast, in low $\epsff$ case, because of the broad range of cell density in star-forming regions, SN explosions in different locations trigger superbubbles with dramatically different sizes.

\subsection{Surface density profiles of molecular gas }\label{sec:results-profile}
\begin{figure}
\includegraphics[width=\columnwidth]{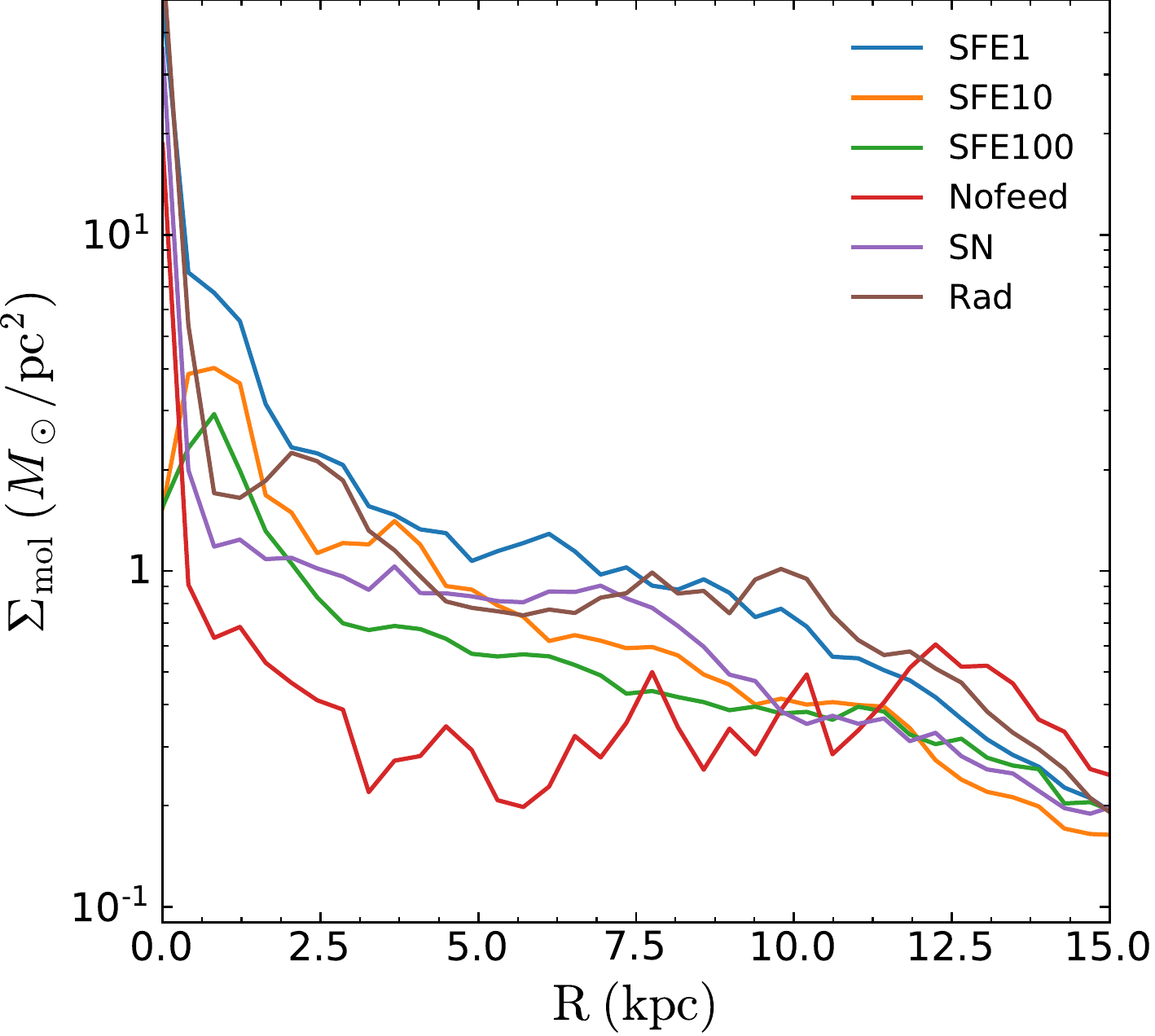}
\vspace{0mm}
\caption{Surface density profiles of molecular gas for the central 15~kpc for different runs. Each line represents the mean profile across many snapshots between 0.4~Gyr and 1~Gyr. The variance around the median profiles is around 0.2--0.5~dex and we do not show this variance for the clarity of the figure. }
  \label{fig:mol-prof}
\end{figure}

We now analyze the molecular content of the galaxy. \autoref{fig:mol-prof} shows the molecular gas surface density profiles of the gaseous disk from different model variations. The profiles are centered at the center-of-mass of the gas disk and extend to 15~kpc. For all model variations, the molecular gas surface density is between $10^5$ and a few times $10^7\Msun/\,\rm pc^{2}$ and decreases with increasing galactocentric radii, consistent with observations.
In the Nofeed run, because of the lack of stellar feedback, the initial star formation burst consumes a large fraction of gas mass during the first few hundred Myr, and leads to a deficit of both atomic and molecular gas after 0.4~Gyr. Therefore, the profile for the Nofeed run is systematically lower than other runs. There is another trend that runs with higher $\epsff$ show lower molecular gas density profiles. Interestingly, the total gas masses in SFE1, SFE10, and SFE100 runs are very similar. Therefore, the lower density profiles in higher $\epsff$ runs is not due to the lack of total gas mass, but the faster consumption of molecular gas by star formation.

\begin{figure}
\includegraphics[width=\columnwidth]{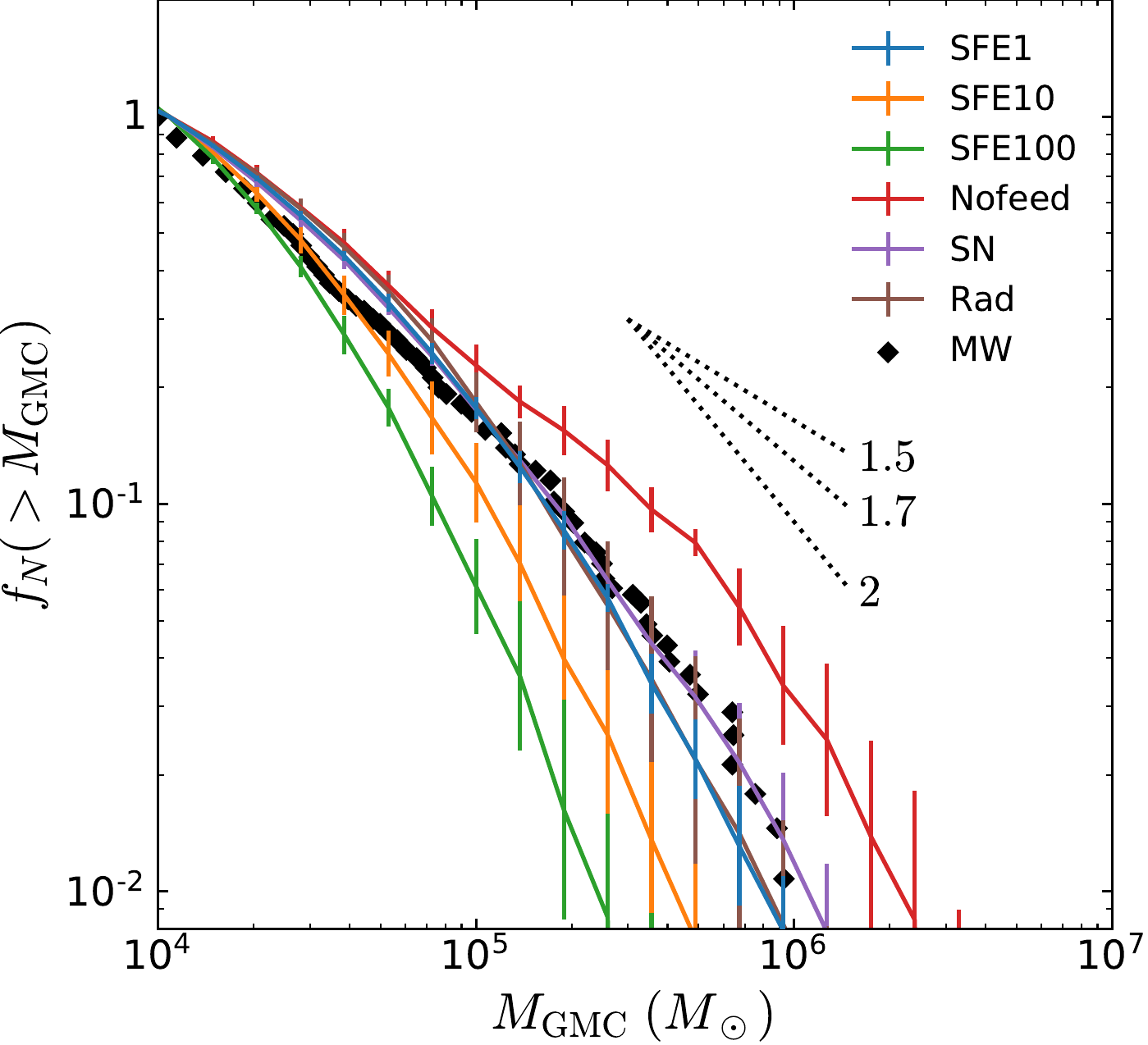}
\vspace{0mm}
\caption{Cumulative mass function of the identified GMCs in different runs. Each solid line shows the median value of the mass function while the errorbars enclose its range across many snapshots from 0.4~Gyr to 1~Gyr. The mass function of Galactic GMCs \citep{fukui_kawamura10} is overplotted as black diamonds for comparison. Here we only include GMCs with mass higher than $10^4\Msun$, therefore all cumulative mass functions are normalized at $10^4\Msun$. Mass functions of different slopes, 1.5, 1.7, and 2, are overplotted as dotted lines for reference.}
  \label{fig:massfunc-runs}
\end{figure}

\subsection{GMC mass function}\label{sec:results-massfunction}

Following \autoref{sec:methods-dendro}, we identify GMCs from many simulation snapshots between 0.4~Gyr and 1.0~Gyr. The identified GMCs span a wide range of mass from $\sim10^7\Msun$ all the way down to the resolution limit of the simulations around several $10^3\Msun$. \autoref{fig:massfunc-runs} summarizes the cumulative mass functions of the model GMCs for all runs. As discussed in \autoref{sec:methods-dendro}, the shape of the mass function at the low-mass end (<$10^4\Msun$) is sensitive to the choice of {\sc astrodendro} parameters. So here we only show results above this mass limit. \autoref{fig:massfunc-runs} also shows the scatter of the mass function for different snapshots over time. We find that the scatter for all runs is relatively small, indicating that the GMC population is in a steady state. This means that the statistically properties of these model GMCs are meaningful and do not depend strongly on the specific epoch of the simulations.

In general, we find that the mass function of model GMCs can be described by a power-law,
\begin{equation}\label{eq:massfunction}
    \frac{dN}{dM}\propto M^{-\beta},
\end{equation}
where $\beta$ is the power-law slope. We fit the mass function with \autoref{eq:massfunction} and obtain the slopes using the maximum likelihood estimation method for all GMC catalogs at different snapshots. The mean and standard deviation of the slopes for each run are listed in \autoref{tab:pl-slope}.

\begin{table}
\centering
 \begin{tabular}{c c c c} 
 \hline
    Name  & $\beta$ & $\beta_{\rm inner}$ & $\beta_{\rm outer}$  \\
 \hline
    SFE1    & $1.78\pm0.01$ & $1.54\pm0.02$ & $1.95\pm0.01$\\
    SFE10   & $1.95\pm0.04$ & $1.81\pm0.07$ & $2.01\pm0.03$\\
    SFE100  & $2.08\pm0.04$ & $1.94\pm0.07$ & $2.14\pm0.04$\\
    Nofeed  & $1.67\pm0.02$ & $1.45\pm0.06$ & $1.70\pm0.02$\\
    SN      & $1.78\pm0.02$ & $1.56\pm0.05$ & $1.93\pm0.03$\\
    Rad     & $1.77\pm0.04$ & $1.58\pm0.03$ & $1.91\pm0.05$\\
 \hline
 \end{tabular}
 \caption{Best-fit power-law slopes of the GMC mass function in six model variations. $\beta$, $\beta_{\rm inner}$, and $\beta_{\rm outer}$ are the mean slopes for the overall GMC sample, GMCs in the inner disk (<5~kpc), and GMCs in the outer disk (>5~kpc), respectively. The standard deviations of the slopes derived from all analyzed snapshots are also listed here for reference.}\label{tab:pl-slope}
\end{table}

It is clear that, in the Nofeed run, the mass function is systematically shallower and extends to larger masses than all other runs with stellar feedback. Moreover, it also shows a statistically significant exponential cutoff at high-mass end. Considering this cutoff and fitting the mass function with a Schechter function, the power-law slope is even shallower than the one that is obtained from pure power-law fit. As we discussed before, because of the lack of stellar feedback to terminate gas accretion onto GMCs, a large fraction of molecular mass is concentrated on a few very massive GMCs, producing an excess number of high-mass GMCs.

On the other hand, the mass functions of SFE1, Rad, and SN runs are not so different from each other and are all very similar to the observed GMC mass function in our Galaxy. The SN run extends to a slightly higher GMC masses, although the difference is within the uncertainty of the time variation during the course of the simulations.
Interestingly, we find a systematic trend that higher $\epsff$ leads to steeper slope of the mass function and smaller maximum GMC mass. These trends are caused by the fast gas consumption and short molecular gas lifetime in high $\epsff$, also seen in \autoref{sec:results-profile}.

\subsection{GMC mass function at different galactocentric radii}\label{sec:results-massfunction-rad}

Recent observations of the Milky Way and nearby galaxies have revealed a systematic variation of the shape of the mass function for GMCs from different locations in the galaxies. GMCs that reside in the inner gas disk tend to have shallower mass function than the ones in the outer disk \citep[e.g.][]{rosolowsky05,rice_etal16}.
Here we explore the spatial variation of the GMC populations in our simulations.

\begin{figure}
\includegraphics[width=\columnwidth]{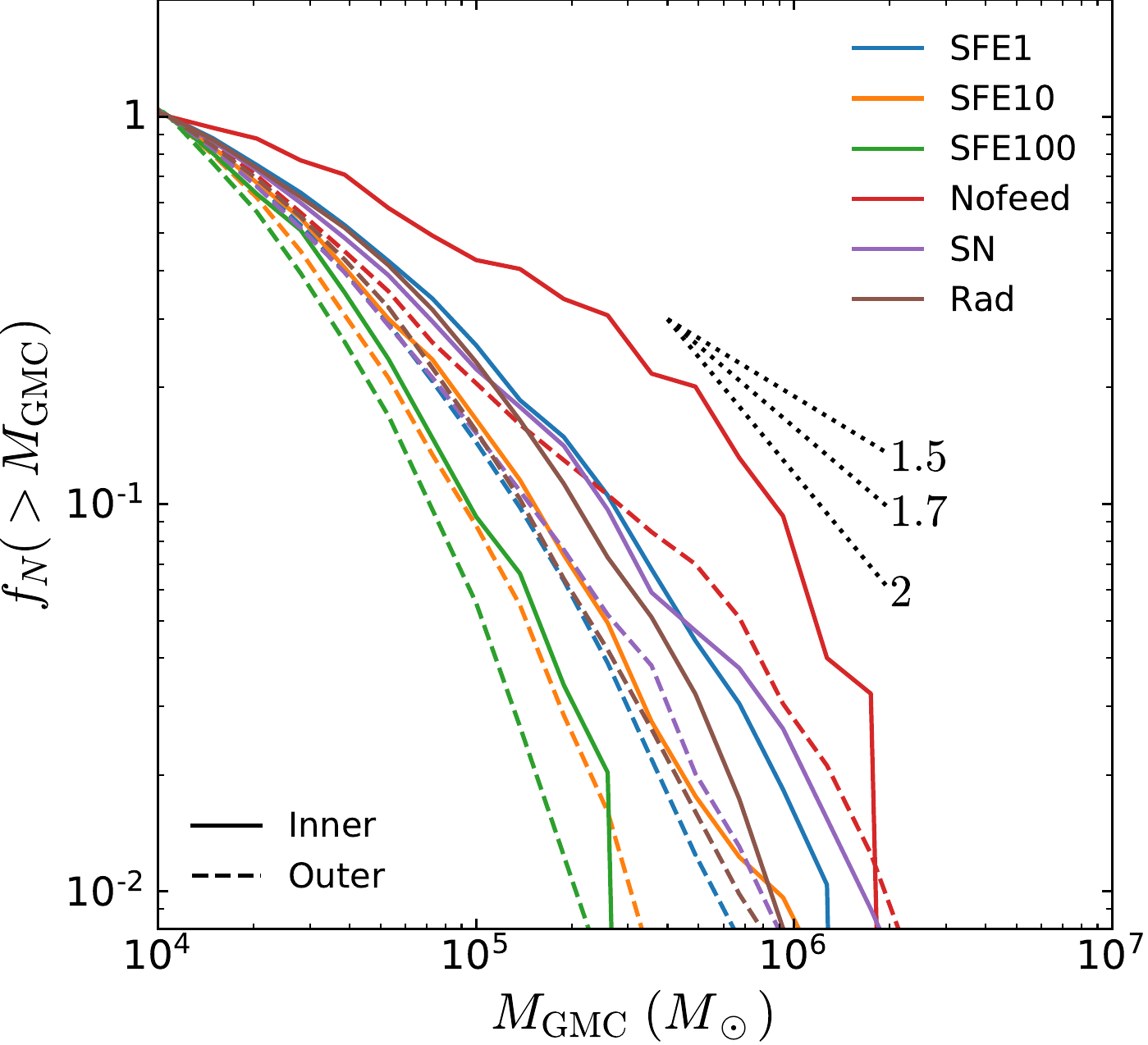}
\vspace{0mm}
\caption{Same as \autoref{fig:massfunc-runs}, but for GMCs in two groups separated by their galactocentric radii: inner (<5~kpc, solid) and outer (>5~kpc, dashed).}
  \label{fig:massfunc-rad}
\end{figure}

We split the whole GMC sample into two groups, inner ($<5$~kpc) and outer ($>5$~kpc), based on their galactocentric radii.
\autoref{fig:massfunc-rad} shows the median of the mass functions of the two groups from many snapshots for all runs. Although the shape of the mass function varies in different model variations as discussed in \autoref{sec:results-massfunction}, all models show the same trend that GMCs in the inner disk have shallower mass functions than that of the outer ones, consistent with observations. We tried different galacto-centric radii from 3-8~kpc to split the GMC sample and found that the trend is robust to choice of this value in this range. For example, in SFE1 run, the slope is 1.54 for the inner group of GMCs and 1.95 for the outer group. These slopes are also quantitatively consistent with the observations of our Milky Way \citep{rosolowsky05}. The mean and standard deviation of the power-law slopes for inner and outer samples for all model variations are listed in \autoref{tab:pl-slope}. Moreover, we find that the mass functions for inner GMCs show a clear cutoff at high mass end, while the outer ones are best described by pure power-law without a statistically significant cutoff \citep[e.g.][]{rice_etal16}. The difference in the mass function between inner and outer disk suggests a strong environmental effects on gas fragmentation by a combination of galactic shear and external pressure, see also \citet{ward_etal16, jeffreson_kruijssen18}.

\subsection{Shape of GMCs}\label{sec:results-shape}

\begin{figure}
\includegraphics[width=\columnwidth]{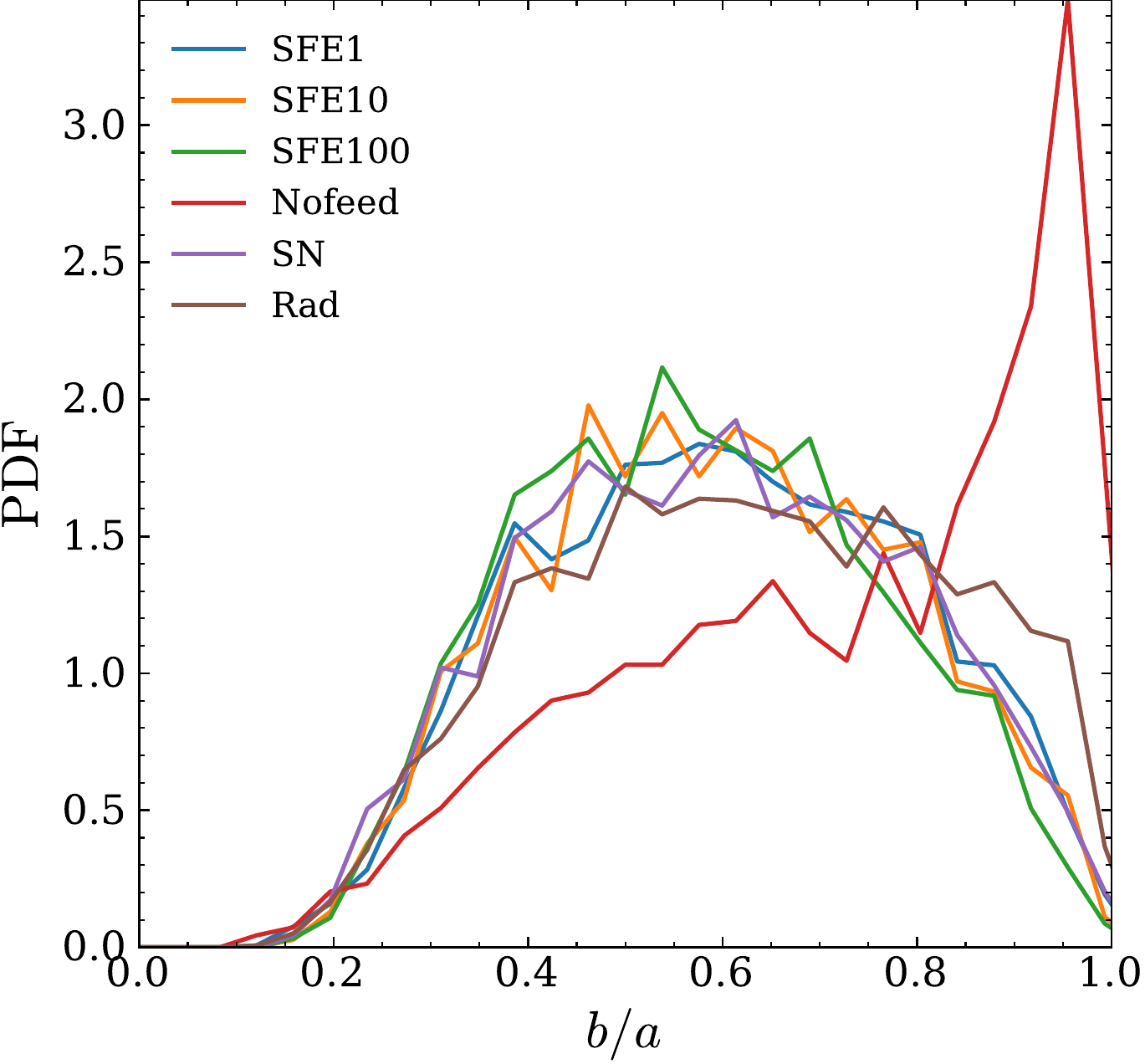}
\vspace{0mm}
\caption{Probability density function of the GMC shape indicator, the ratio between semi-minor and semi-major axis, for all GMCs with mass higher than $10^4\Msun$ in different runs. The same as previous figures, each solid line shows the median value of the distribution across many snapshots from 0.4~Gyr to 1~Gyr. We do not show the variance around each line in the figure for clarity as the variance is very small, $<0.1$~dex.
}
  \label{fig:gmc-shape}
\end{figure}

As discussed in \autoref{sec:methods-dendro}, {\sc astrodendro} identifies GMCs based on the isodensity contours of the molecular map and organizes the GMC sample in a hierarchical way. After the envelop of an identified GMC is determined, {\sc astrodendro} also models the best-fit ellipses, which can be used to study its intrinsic shape in 2D. In \autoref{fig:gmc-shape}, we show the distribution of the ratio between semi-minor ($b$) and semi-major axis ($a$) of all GMCs larger than $10^4\Msun$ in different runs. We find that stellar feedback not only changes the baryon cycle of star-forming regions but also reshapes the morphology of individual GMCs. In Nofeed run, the shape distribution peaks towards $b/a\sim1$, suggesting that a large fraction of the GMCs are spherical. However, once stellar feedback is included, even in Rad or SN runs, the shape of GMCs deviates from spherical with a distribution of $b/a$ peaks around 0.5. Revisiting the right panel of \autoref{fig:prj}, we can see that the most of the GMCs are distributed along the filamentary structure of the gas disk. The direction of the semi-major axis is mostly parallel to the orientation of the filaments. The existence of the filamentary structures is due to the feedback-induced superbubbles that constantly compress the gas around different star-forming regions. GMCs in Rad run shows a slightly more spherical shape than those in other runs with SN feedback, possibly because SN feedback creates bubble-like structures more easily in the gas disk than the radiative feedback. To better quantify the difference of the distribution of GMC shapes in different runs, we calculate the Kolmogorov-Smirnov statistics and obtain the p-values for each pair of model variations. The p-values between the distributions in SFE1, SFE10, and SN are all above 0.1 suggesting that the GMC shape is statistically indistinguishable. The GMC shape in SFE100, although similar by eye, shows significant difference compared to those in other SFE runs, with p-values $1.26\times10^{-5}$ (vs. SFE1) and 0.007 (vs. SFE10). Finally, the GMC shape in Nofeed, Rad runs are dramatically different with respect to all other runs, with p-values almost zero.

\subsection{Two-point correlation function of the GMC spatial distribution}\label{sec:results-TPCF}
The spatial distribution of GMCs reflects the degree of clustering and dynamical interaction among different star-forming regions across the whole gaseous disk.
One of the most common spatial statistics that characterizes the spatial correlations across various length scales is the TPCF. To compare with recently observations, such as \citet{grasha_etal18}, here we only consider the TPCF in 2D.

\begin{figure}
\includegraphics[width=\columnwidth]{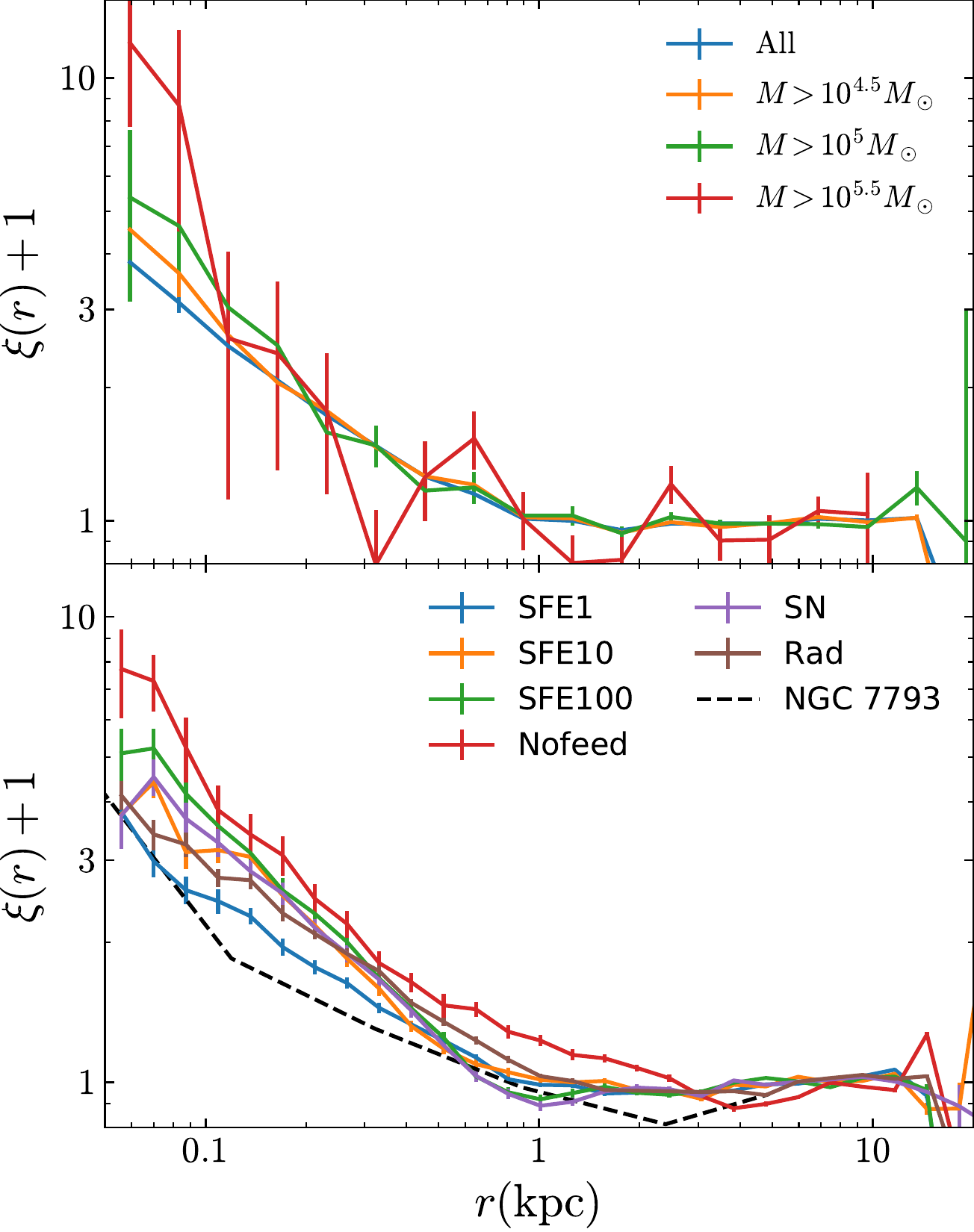}
\vspace{0mm}
\caption{\textit{Top:} TPCF of the GMCs in different mass ranges in SFE1 run. Blue line shows the TPCF of the whole GMC sample ($>10^4\Msun$), while yellow, green, and red lines show the GMCs in different mass bins, as described in the legend. More massive GMCs are more strongly clustered on smaller scales, a trend that exists in all our feedback runs. \textit{Bottom:} TPCF of the whole GMC samples in all six runs. As in \autoref{fig:massfunc-runs}, each solid line shows the median value of the TPCF while the errorbars enclose its range across many snapshots from 0.4~Gyr to 1~Gyr. For comparison, we overplot the observed TPCF for GMCs identified from NGC 7793 \citep{grasha_etal18}.
}
  \label{fig:TPCF-runs}
\end{figure}

We calculate the TPCF from the 2D position of the identified GMCs over the scales between 0.05 and 20~kpc. The upper panel of \autoref{fig:TPCF-runs} shows the TPCF of the GMCs in the SFE1 run. We find that the TPCF decreases with increasing correlation length: the GMC distribution shows a strong clustering on 0.1~kpc scales and almost no clustering on scales larger than 1~kpc. Moreover, we split the GMC sample into different mass bins and investigate how the TPCF changes with GMC masses. We find that more massive GMCs tend to be more clustered compared to less massive ones on scales below 0.2~kpc. This mass segregation is consistent with the recent spatial analysis of the GMC distribution in NGC 7793 \citep{grasha_etal18} and M51 \citep{grasha_etal19}. We repeat the same procedure for all model variations and found that this trend exists for all runs, suggesting that the mass-dependent clustering is a general feature of hierarchical fragmentation.

On the lower panel, we systematically study the TPCF of GMCs in all model variations. We find a distinct difference of the TPCF for different models. Nofeed run shows the strongest correlation among all models over a wide range of length scales below 3~kpc. This suggests that fragmentation of the cold gas leads to a strong clustering of GMC distribution, and, without stellar feedback, the clustered cold gas is unable to be dispersed to reduce the level of correlation.
With stronger feedback, the amount of clustering is reduced as star-forming regions are continuously disrupted on a timescale much shorter than the dynamical time of the gas disk. Cold gas that is not forming stars is quickly recycled back to the ISM and is redistributed on different scales depending on how far the effects of stellar feedback may reach out. 

We note that even in Rad or SN runs where only a subset of stellar feedback mechanisms is turned on, the level of correlation is notably reduced compared to the Nofeed run. Interestingly, Rad run shows stronger correlations on scales $>0.2$~kpc but weaker correlations for $<0.2$~kpc than the SN run. This result is consistent with our physical intuition that radiative feedback can disperse cold gas on scales of star-forming regions but cannot easily affect the gas distribution over kpc scales, while SN feedback can generate large-scale superbubble but cannot reduce the small scale clustering because the dynamical time of the star-forming regions are usually shorter than the lifetime of the massive stars for SN explosions. As expected, when both the radiative and SN feedback channels are included (SFE1), the correlation is reduced simultaneously in both scales.

Interestingly, we find that runs with higher $\epsff$ show stronger correlation than the lower $\epsff$ ones, see the results for SFE1, SFE10, and SFE100 runs on the lower panel of \autoref{fig:TPCF-runs}. As we discussed in \autoref{sec:results-prj}, higher $\epsff$ leads to smaller dynamical range of gas density in star-forming regions as the densest gas is consumed more quickly. Therefore, stellar feedback launched from these regions affect clouds with densities in a narrower range than feedback in the low $\epsff$ case, explaining the better-ordered bubbles of similar sizes in high $\epsff$ runs. GMCs are then distributed along the edge of the overlapping bubbles of similar sizes, which leads to a higher level of clustering on scales below the bubble size (<0.5~kpc, scale length of the gas disk).

The sensitivity of the TPCF on the choice of different subgrid models makes it a powerful observable to calibrate these subgrid models. Here we compare our results to the recent ALMA-LEGUS observations. \citet{grasha_etal18} obtained a relatively weak correlation for the spatial distribution of GMCs in NGC 7793, which can only be reproduced by our SFE1 run, albeit with a slight overestimation of the correlation on scales $<0.5$~kpc. We hasten to add that, although insightful, the comparison with NGC 7793 cannot be used to directly assess whether SFE1 is the best run among all models. Besides the fact that these measurements correspond to a single only galaxy, NGC 7793 has also a different size and mass compared to our Galaxy. The changes on TPCF for different types of galaxies can be dramatic. Another galaxy in ALMA-LEGUS, M51, shows an extremely flat TPCF with almost no correlation across all scales \citep{grasha_etal19}. 

\section{Discussion}\label{sec:discussion}

\subsection{Feedback-regulated star formation activities}\label{sec:discussion-SFH}

In \autoref{sec:results-SFH}, we find that the Rad run with only radiative feedback shows a systematically higher SFR compared to all other runs that have SN feedback throughout the course of the simulations. As demonstrated in M19, momentum injection from radiative feedback is only 5--10\% of the one from SNe. The overall momentum feedback budget and the ability of generating large-scale galactic winds are largely dominated by the SN feedback.  Still, the SFR in Rad run is not dramatically different from other feedback runs, suggesting radiative feedback alone is also able to keep the SFR within a reasonable range.
We emphasize that the similarity of the SFR in all feedback runs is only applicable to the current initial condition of a Milky Way-sized galaxy with relatively low gas fraction. In galaxies with higher gas fraction or more compact gas distribution, which are commonly seen in high-$z$ galaxies, the star formation activity can be dramatically different when using different feedback channels.

\subsection{Comparison to previous simulations}\label{sec:discussion-sims}

During the last decade, a number of studies have investigated the effects of star formation and stellar feedback models on the structure of the ISM using different numerical simulations.
\citet{hopkins_etal12} performed a suite of isolated galaxy simulations with various stellar feedback prescriptions. They did not find a dramatic difference of the mass function when switching on and off radiative or SN feedback, consistent with our result. However, the mass function from all of their runs extends to much higher masses than that in our simulations and in observations.
The difference may be caused by the use of different GMC identification methods. \citet{hopkins_etal12} used SUBFIND, a halo finder commonly used in galaxy formation simulations, to identify the bound overdensities of gas particles. This approach can only be applied to numerical simulations and cannot be used in observations, which do not have the full 6D information.
In contrast, we adopt the \textsc{astrodendro} algorithm described in \autoref{sec:methods-dendro} and try to mimic the observational approach and identifies GMCs from 2D molecular gas projections.
As shown in \citet{grisdale_etal18}, the properties of GMCs vary dramatically depending on whether they are identified in 2D or 3D.
In terms of the spatial distribution of neutral gas, \citet{grisdale_etal17} found that the density power spectra of HI gas is much higher in the no feedback run than that in full feedback runs \citep[see also][]{combes_etal12, walker_etal14}, because stellar feedback injects turbulent kinetic energy and reduces the strength of correlation below the scales of $\sim1$~kpc. In \autoref{sec:results-TPCF}, we find that stellar feedback also affects the distribution of dense molecular gas on a similar fashion and reduces the correlation on similar scales ($<1$~kpc).

In \autoref{sec:results}, we showed that both the mass function and TPCF of the model GMCs depend strongly on the choice of $\epsff$, suggesting that these observables may be used to constrain the appropriate value of $\epsff$. In fact, similar efforts have been made from our previous work, where we used the properties of young star clusters to constrain $\epsff$ \citep{li_etal18}. Based on a novel implementation that treats star cluster as a unit of star formation in cosmological simulations \citep{li_etal17}, we found that a high efficiency ($\epsff>0.1$) is needed to reproduce the observed correlation between star formation rate surface density and star cluster formation efficiency. This conclusion seems in contradiction to the results we show in the current work, where we find that $\epsff=0.01$ best reproduces the mass function and TPCF of GMCs. However, we emphasize here that \citet{li_etal18} performed the test of $\epsff$ in high-$z$ galaxies, which is typically much more gas-rich and turbulent-supported \citep{meng_etal19} than the $L_*$ galaxies on the star-formation main sequence at $z=0$. It is possible that the efficiency changes with the physical condition of the gas components in different types of galaxies. Actually, several numerical explorations suggest a variable efficiency that depends on both the Alfv\'en and sonic Mach number \citep[e.g.][]{padoan_etal12, federrath_klessen12, semenov_etal16}. Exploring this environment-dependent star formation efficiency, however, is beyond the scope of this paper.

\subsection{Caveats}
We note that there are a few caveats in the simulations as well as the analysis methods. First, for all simulations presented in this work, the radiative feedback, such as photoheating and radiation pressure, is not modeled with direct radiative transfer but is based on an effective subgrid model. Therefore, the effects of radiative feedback can be either over- or under-estimated.
Second, as the simulations do not follow the chemical network of molecular hydrogen formation and destruction, the distribution of $\rm H_2$ is calculated based on a subgrid model described in \citet{mckee_krumholz10}. In realistic astrophysical environments, the molecular fraction depends strongly on the local radiation fields, which are largely affected by the adjacent star-forming regions. This missing physics in our current model can lead to inaccurate results, especially for molecular gas around young star clusters.
A self-consistent radiation hydrodynamics simulation \citep[e.g.][]{wise_etal12,rosdahl_etal15,li_etal17,kannan_etal19} is needed to accurately estimate the mass budget of molecular gas in the galaxy. Lastly, in this work, we focused on studying an isolated $L_*$ galaxy at $z=0$. Therefore, the conclusions are mostly limited to this type of galaxies. Future works on simulating different types of galaxies, such as high-$z$ gas-rich analogs, merging systems and low mass dwarfs, are needed to investigate the molecular gas properties on a wide range of galaxy types (Li et al. in preparation).

\section{Summary}\label{sec:summary}

We performed a suite of isolated galaxy disk simulations with different subgrid models and parameters under the framework of the {\sc Arepo/SMUGGLE} model. We identified GMC candidates using the hierarchical clump-finding algorithm, {\sc astrodendro}, and investigated various properties of the GMC populations, such as mass, shape, and spatial correlations, in six model variations. We found that, although the SFH of the galaxy is not sensitive to the choice of different subgrid models (except for the no feedback case), the properties of the identified GMCs vary significantly. Below we list our main findings.

\begin{itemize}
    \item The star formation rate of the simulated MW-sized galaxies is around $1$--$10\Msun/$yr for runs with stellar feedback, while for no feedback run, the rate reaches $>30\,\Msun/$yr during the first few dynamical times. We also notice that, even for runs with only radiative or SN feedback alone, the star formation activities can still be regulated to a similar level as the full feedback case.
    \item We identify the GMC populations from the 2D molecular gas surface density distribution. In general, the mass function of GMCs can be described as a power-law. The power-law slopes for the full feedback, radiative-only, and SN-only runs with $\epsff=0.01$ are around $1.78\pm0.01$, consistent with observations. However, without feedback, this slope is systematically shallower and the mass function extends to much higher masses.
    \item Runs with higher $\epsff$ tend to have less amount of molecular gas reservoir because of faster gas consumption rate due to star formation. This trend in turn affects the shape of the GMC mass function: runs with higher $\epsff$ have steeper power-law slopes. In the extreme case when $\epsff=1.0$ is adopted, the slope is as steep as $2.08\pm0.04$.
    \item The GMC mass function shows a systematically shallower power-law slope for GMCs in the inner part of the galaxy than those in the outer part, a trend that appears on all model variations. In particular, the simulations with $\epsff=0.01$ reproduce the observed slopes of mass function for both inner and outer part of galaxies.
    \item We obtain the best-fit ellipses for each individual GMCs by estimating the first and second moments of the structure with {\sc astrodendro}. We find that most of the GMCs are highly non-spherical in all feedback runs, with a distribution of the ratio between semi-minor and semi-major axis peaked around 0.5. In contrast, GMCs in the no feedback run are mostly spherical.
    \item We calculate the TPCF of the spatial distribution of the GMCs. In general, GMCs have stronger spatial correlations on smaller scales. Moreover, we find a clear trend indicating that GMCs with higher masses are more correlated than the lower mass ones on scales smaller than $0.3$~kpc. Both trends exist in all model variations and are qualitatively consistent with observations.
    \item Without feedback, the TPCF is systematically higher than all feedback runs. Stellar feedback destroys individual star-forming regions, redistribute molecular gas across the galactic disk, and reduces the correlation strength. Interestingly, different stellar feedback mechanisms control the GMC correlation on difference scales. With radiative feedback, the TPCF is reduced most significantly on scales below 0.2~kpc, while with SN feedback on scales larger than 0.2~kpc.
    \item Moreover, higher $\epsff$ leads to higher TPCF on scales $<1$~kpc since higher $\epsff$ helps remove dense gas faster and facilitates supernovae blowing out well-organized bubbles of similar sizes more easily.
\end{itemize}

Overall, we highlight that the properties of GMCs are strongly affected by different choices of subgrid models. By comparing the simulation results, such as mass functions and TPCF, to sub-mm observations of nearby galaxies, the subgrid models can be better constrained. Future observations of molecular gas distributions in nearby galaxies will provide us with great opportunities to improve galaxy formation simulations that are aimed to resolve the star-forming regions and multi-phase ISM.

\section*{Acknowledgements}
We thank the anonymous referee for detailed comments and suggestions that help improve the manuscript. We are grateful to Oleg Gnedin, Shanghuo Li, Oscar Agertz, Florent Renaud, Rahul Kannan, Aaron Smith, Vadim Semenov, Nick Gnedin, Greg Bryan, Miao Li, Drummond Fielding, and Diederik Kruijssen for useful discussion and insightful comments. We would like to thank Kathryn Grasha and Viviana Casasola for kindly providing us the molecular data for comparison with our simulations. This research made use of {\sc astrodendro}, a Python package to compute dendrograms of Astronomical data (\url{http://www.dendrograms.org/}).
HL was supported by NASA through the NASA Hubble Fellowship grant HST-HF2-51438.001-A awarded by the Space Telescope Science Institute, which is operated by the Association of Universities for Research in Astronomy, Incorporated, under NASA contract NAS5-26555. MV acknowledges support through an MIT RSC award, a Kavli Research Investment Fund, NASA ATP grant NNX17AG29G, and NSF grants AST-1814053 and AST-1814259. FM is supported by the Program ``Rita Levi Montalcini'' of the Italian MIUR.  LVS is
thankful for financial support from the Hellman Foundation as
well as NSF and NASA grants, AST-1817233 and HST-AR-14552. The simulations of this work were run on the Harvard Odyssey clusters and the Comet HPC resource at San Diego Supercomputer Center as part of XSEDE through TG-AST180025.

\appendix

\section{GMC mass functions under different \textsc{astrodendro} parameters}\label{sec:appendix}

\begin{figure*}
\includegraphics[width=2\columnwidth]{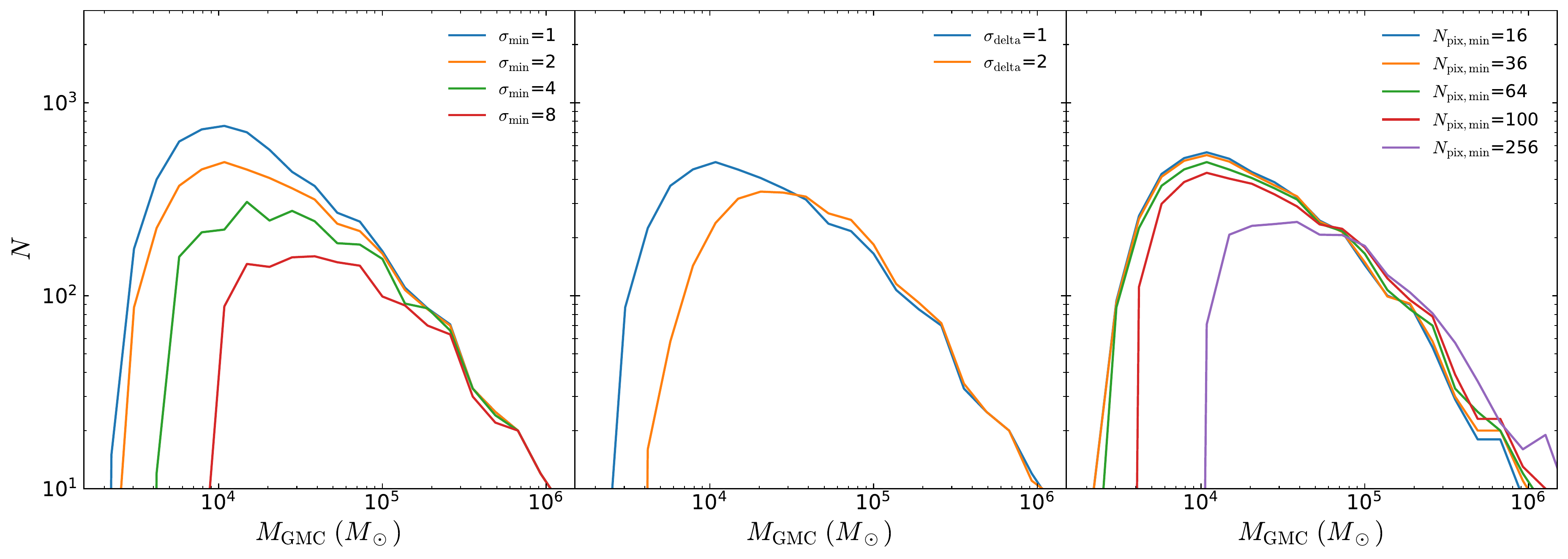}
\vspace{0mm}
\caption{Mass function of identified clumps using different parameters described in \autoref{sec:methods-dendro}. Varying three parameters, $\sigma_{\rm base, min}$ (left), $\sigma_{\rm delta, min}$ (mid), and $N_{\rm pix, min}$ (right), results in some differences in clump mass function. Fortunately, all three parameters affect mostly the mass distribution of clumps in low mass end. The high mass clumps with mass larger than $2\times10^4\Msun$ are not largely affected by the choice of these parameters over a wide range.}
  \label{fig:dendro-para}
\end{figure*}

As described in \autoref{sec:methods-dendro}, for all GMC samples identified in the main text, we use the \textsc{astrodendro} parameters that best mimic the recent ALMA observations. The fiducial values of the three parameters are $\sigma_{\rm min}=2\sigma_{\rm ALMA} $, $\sigma_{\rm delta}=\sigma_{\rm ALMA}$, and $N_{\rm pix,min}=64$. To test the sensitivity of the choice of \textsc{astrodendro} parameters on cloud identification, we show how mass function changes as we vary these three parameters around the fiducial values in \autoref{fig:dendro-para}.

We find that higher $\sigma_{\rm min}$ suppresses the number of identified low mass clouds. With an extreme value $\sigma_{\rm min}=8$, almost no cloud less massive than $10^4\Msun$ is identified. This is because 1) less massive clouds whose surface density are below this threshold are completely removed from the sample 2) the boundary of massive clouds shrinks and therefore the total mass of the identified clouds is less than that is estimated in lower $\sigma_{\rm min}$ case. Similar behavior is also seen when changing $N_{\rm pix, min}$. 
Moreover, the smaller the value of $\sigma_{\rm delta}$, the easier larger GMC complexes are split into smaller pieces. Therefore, the mass function with $\sigma_{\rm delta}=1$ extends more at the lower mass end while shows a slight deficit around $10^5\Msun$ compared to the $\sigma_{\rm delta}=2$ case. Despite all the above chances, the high mass end of the mass function, i.e. $M>2\times10^4\Msun$ does not change much as we vary these parameters over a reasonable range, suggesting that the identified GMCs are robust to the choice of these \textsc{astrodendro} parameters when we focus on massive GMCs that are well resolved by the gas cells.

\section*{Data availability}

The data that support the findings of this study are available from the corresponding author, upon reasonable request.

\bibliographystyle{mnras}
\bibliography{references} 

\bsp	
\label{lastpage}
\end{document}